\documentclass[pdf, twocolumn,superscriptaddress,floatfix,showpacs,amsmath,longbibliography,nofootinbib,amssymb]{revtex4-1}
\usepackage[colorlinks=true,citecolor=blue,linkcolor=blue]{hyperref}
\usepackage[usenames]{color}
\usepackage{subfigure}
\usepackage{dcolumn}
\usepackage{mathrsfs}
\usepackage{bm}
\usepackage{amsmath,amssymb,epsfig,float,graphics}
\usepackage{appendix}
\newcommand{\bk}{\mathbf{k}}

\newcommand{\colb}[1]{\textcolor{blue}{#1}}

\begin{document}
\baselineskip=0.45 cm

\title{Probing Lorentz-invariance-violation with quantum coherence of Unruh-DeWitt detector}

\author{Yihao Wu}
\affiliation{School of Physics, Hangzhou Normal University, Hangzhou, Zhejiang 311121, China}

\author{Xiaobao Liu}
\affiliation{Department of physics and electrical engineering, Liupanshui Normal University, Liupanshui 553004, Guizhou, China}

\author{Zehua Tian}
\email{tzh@hznu.edu.cn}
\affiliation{School of Physics, Hangzhou Normal University, Hangzhou, Zhejiang 311121, China}

\begin{abstract}
Testing Lorentz invariance violation (LIV) is notoriously difficult, as its characteristic energy scale, $M_\star$, in modified dispersion relations typically lies beyond current experimental reach. To enable a low-energy probe, we investigate the quantum coherence of an inertial Unruh-DeWitt (UDW) detector interacting with a massless scalar field with a Lorentz-violating dispersion relation $\omega_{|\textbf{k}|}=|\textbf{k}|f(|\textbf{k}|/M_\star)$. 
Unlike the Lorentz-invariant case where quantum coherence is rapidity-independent, we find for the Lorentz-violating quantum field case
the dynamics of detector's quantum coherence exhibits a strong dependence on its rapidity---offering a potential low-energy signature of LIV.
Applying this to the polymer-quantized scalar field theory inspired by loop quantum gravity,
we show that the detector's quantum coherence exhibits a pronounced dependence on rapidity and undergoes a sharp transition near a critical rapidity $\beta_c \approx 1.3675$,  a value within reach of existing facilities like the Relativistic Heavy Ion Collider. We also show how the detector's rapidity and its energy-level spacing enhance the response to the test of LIV. 
Our results establish quantum coherence as a sensitive and practical probe of LIV, providing a complementary avenue for testing quantum gravity-induced modifications to field theory.
\end{abstract}

\baselineskip=0.45 cm
\maketitle
\newpage

\section{Introduction}
Lorentz invariance lies at the heart of modern physics, underpinning both special relativity and the standard model of particle physics~\cite{Mattingly2005,Amelino2013}. Its validity forms the foundation for our understanding of spacetime symmetries and field dynamics. Nevertheless, a wide class of quantum gravity theories, such as string theory~\cite{S. Samuel}, brane-world scenarios~\cite{Randall1999,Burgess2002}, and loop quantum gravity~\cite{R. Gambini,Rovelli2007,Thiemann2008}, predict that Lorentz symmetry may be only approximate, potentially breaking down at sufficiently high energies close to the Planck scale.
Detecting even a slightest deviation from exact Lorentz symmetry would therefore offer a crucial window into Planck-scale physics, and provide empirical guidance for the unification of quantum mechanics and general relativity.
Experimental searches for LIV have been carried out across a broad range of energy scales, from $\gamma$-ray dispersion~\cite{AmelinoCamelia1998,Jacob2008,Abdo2009}, vacuum birefringence~\cite{KosteleckyMewes2002,Laurent2011,Toma2012} and modifications in particle decay thresholds~\cite{Coleman1999,Jacobson2003}, to canonical commutation relation of the centre-of-mass mode of a mechanical oscillator \cite{Igor, Kumar}. These tests have yielded stringent upper bounds on LIV parameters, yet the expected effects are typically suppressed by powers of the ratio $(E/M_\star)^n$, where $M_\star$ denotes the LIV energy scale, often assumed to be near the Planck energy ($\sim10^{19}\rm GeV$). Consequently, current accelerator and astrophysical observations, spanning from $\rm TeV$ collisions to ultra-high-energy cosmic rays  ($\sim10^{11}\rm GeV$), remain many orders of magnitude below the scale at which such violations are theoretically expected to emerge~\cite{M. Takeda}. While these high-energy observations have constrained the phenomenology of Lorentz violation, 
improving the energy scale to the Planck energy to probe the Lorentz violation remains challenging. Therefore, an open problem arises: is it possible for the effective low-energy theory to reveal imprints of the theory's high-energy structure?

From the perspective of quantum field theory, even strong Planck-scale violations can manifest as subtle low-energy traces through effective mechanisms, such as modified vacuum expectation values or altered field dispersion relations~\cite{Mattingly2005}. This insight has motivated the search for indirect low-energy probes capable of detecting Lorentz violation through its influence on quantum fields~\cite{J. Collins, J. Polchinski}. One of the most promising frameworks for such studies is provided by the UDW detector~\cite{Unruh1976,DeWitt1979,Birrell1982,Takagi1986,Crispino2008,Hu2012}, i.e., a two-level quantum system interacting with a quantum field. The UDW detector has proven to be a versatile theoretical tool for exploring relativistic quantum information, quantum thermodynamics, and field quantization under noninertial motion~\cite{Louko2008, Louko2006, Lin2017, Eduardo2013, Rick2024, Keith2018, Henderson2018, Tian2023, Yu2005,Zhu2010,Arias2016,Liu2018,Liu2021,Liu2025}. Recent works have shown that in certain Lorentz-violating field models, the excitation rate of inertial detector moving with constant velocity can undergo abrupt changes when exceeding  a critical rapidity, suggesting a potential low-energy signature of high-energy symmetry breaking~\cite{V. Husain, Nirmalya22016, Louko22018, Stargen22017, Nirmalya22018, Tian2025}. In particular, its relevant proof-of-principle demonstration has been proposed using quantum fluid platform \cite{Tian2021, Tian2022, Tian2026}, thus providing an experimentally realizable test field to verify whether the effective low-energy theory can reveal unexpected imprints of the theory’s high-energy structure, in quantum field theory.

On the other hand, quantum coherence, rooted in the superposition principle, is one of the key fundamental aspects of quantum physics ~\cite{Leggett1980}. 
Similar to other quantum resources, such as entanglement and quantum discord, quantum coherence is regarded as an important physical resource for a variety of fields,
ranging from quantum information processing to quantum sensing and metrology, thermodynamics, and biology \cite{Streltsov2017}.
Despite the fundamental importance of quantum coherence, only recently Baumgratz \emph{et al.} have introduced a rigorous framework for the quantification of coherence and identified intuitive and easily computable measures of coherence, such as  the $l_1$ norm~\cite{Baumgratz2014},
\begin{eqnarray}\label{l1}
C_{l_1}=\sum_{i\neq j}|\rho_{i,j}|,
\end{eqnarray}
with $\rho$ being the density matrix of quantum system. Since then, the characterization, quantification, manipulation, dynamical evolution, and operational application of quantum coherence has attracted a lot of attention \cite{Streltsov2017, Hu2018}.
Furthermore, the study of quantum coherence has been extended to the relativistic framework, known as the relativistic quantum information \cite{RQI}.
Wang \emph{et al.} investigated how the relativistic motion degrades quantum coherence of a UDW detector \cite{Wang2016}. Liu \emph{et al.} explored 
how to use the boundary to protect quantum coherence of two-level atoms from vacuum fluctuations of quantum fields \cite{aLiu2016, bLiu2016, Liu2022}. Quantum coherence of quantum field has also been studied in different scenarios, such as the noninertial framework \cite{Saveetha2022, Du2024, Li2024, Wu2021}, de Sitter universe \cite{Wus2023} and black hole background \cite{Wus2024}. The behavior of quantum coherence of UDW detector has also been applied to the detection of different physics, including coupling form between UDW detector and quantum fields \cite{Barros2025}, gravitational waves \cite{Barros2024}, properties of coupled quantum fields \cite{Nicolaos2023, TianSCPMA2023}, and so on. In addition to that, quantum coherence has been studied in the field of 
relativistic quantum thermodynamics, by exploring its effect on relativistic quantum Otto engine \cite{Enrique2018, Finnian2018, Arnab2022, Dimitris2025, Vahid2025, Hao2020, Rudra2025, Nikos2024, Tomoya2025, DimitrisarXiv2025}, and charging of relativistic quantum battery \cite{TianJHEP2025, LiuSCPMA2025, ArnabarXiv2025, YanPRD2025, ZhiAS2025, RahularXiv2025}.

In this work, we propose to employ the quantum coherence of an UDW detector as a novel probe of LIV in quantum field theory. Specifically, we focus on a polymer-quantized scalar field, an effective field model inspired by loop quantum gravity~\cite{Rovelli2007,Thiemann2008, 2Husain2010} that naturally incorporates LIV through a modified dispersion relation. By analyzing the evolution of the detector's coherence when coupled to this field, we demonstrate that the coherence exhibits a pronounced dependence on the detector's rapidity and undergoes a sharp transition near a critical value $\beta_c \approx 1.3675$. This behavior contrasts sharply with the Lorentz-invariant case, in which coherence decay is entirely rapidity-independent. The detector-rapidity-dependent property of quantum coherence and 
the existence of a critical rapidity at which the quantum coherence changes sharply via the detector's rapidity 
provides a clear and potentially measurable signature of LIV. 
Remarkably, the predicted $\beta_c$ falls within the operational range of current accelerator facilities such as the Relativistic Heavy Ion Collider~\cite{velocity}, indicating that this effect could be experimentally accessible in principle.
Our analysis thus establishes quantum coherence as a viable and sensitive probe of LIV in quantum field theory, complementing traditional symmetry tests based on particle dispersion or decay thresholds. It might bridge low-energy quantum information measures with high-energy quantum gravity phenomenology, offering a new avenue to test Planck-scale physics within experimentally reachable regimes.

This paper is organized as follows. In Sec.~\ref{section2}, we introduce the general framework of Lorentz-violating quantum field theory and the evolution of the UDW detector. In Sec.~\ref{section3}, we analyze coherence dynamics in both scenarios: Lorentz-invariant and Lorentz-violating field cases, establishing a general detection criterion. Section~\ref{section4} applies this framework to the polymer-quantized field, exploring parameter dependencies of quantum coherence and identifying its corresponding critical rapidity. Finally, Sec.~\ref{section5} presents the conclusions and discussions.

Throughout, natural units ($c=\hbar=1$) are used.

\section{Lorentz-violating quantum field theory and evolution of UDW detector} \label{section2}
In this section, we will introduce the topic of Lorentz-violating quantum field theory, as well as the model of a two-level UDW detector that is coupled to the quantum field.

\subsection{Lorentz-violating quantum field theory}
Let us start with four-dimensional Minkowski spacetime, whose corresponding line element can be written as
\begin{eqnarray}
ds^2=dt^2-dx^2-dy^2-dz^2.
\end{eqnarray}
In this four-dimensional Minkowski spacetime, we consider a massless scalar field $\Phi(\textbf{x})$ with the Lorentz-violating dispersion relation
\begin{eqnarray}\label{omega}
	\omega_{|\textbf{k}|}=|\textbf{k}|f(|\textbf{k}|/M_\star),
\end{eqnarray}
which denotes the corresponding energy with the spatial momentum $\textbf{k}$. Moreover, the positive constant $M_\star$ dominates the energy scale of Lorentz violation, and $f$ is a smooth positive function on the positive real line. If we define a dimensionless quantity $g=|\textbf{k}|/M_\star$ for convenience,
the Lorentz-invariant dispersion relation should be recovered when $g=|\textbf{k}|/M_\star\ll 1$, i.e., $\omega_{|\textbf{k}|}\approx|\textbf{k}|$. and in such case, the function $f$ satisfies the limit condition, $\lim_{g\rightarrow0_+}f(g)\rightarrow1$.

Decomposing the field operator of the massless scalar field $\Phi(\textbf{x})$ into spatial Fourier modes, a mode with spatial momentum $\bk\neq0$ is a harmonic oscillator with
the angular frequency $\omega_{|\textbf{k}|}$ given in Eq.~(\ref{omega}). In this case, the field operator of the massless scalar field can be written as
\begin{eqnarray}\label{Phi}
\Phi(\textbf{x}) =\int{d^3\textbf{k}\,\rho_{|\textbf{k}|}(a^\dagger_{\textbf{k}}e^{-i\textbf{k}\cdot\textbf{x}}+a_{\textbf{k}}e^{i\textbf{k}\cdot\textbf{x}})},
\end{eqnarray}
where $\textbf{x}=(x, y, z)$, $\rho_{|\textbf{k}|}=d(|\textbf{k}|/M_\star)/\sqrt{(2\pi)^3|\textbf{k}|}$ indicates the density-of-states weight factor with $d$ being a smooth complex-valued function on the positive real line.
Note that if $f(g)=1$ and $|d(g)|=1/\sqrt{2}$,
the field is the Lorentz-invariant massless scalar field.

\subsection{Evolution of UDW detector}
To probe the LIV in quantum field theory, we will employ a two-level atom as the UDW detector that is coupled to the quantum field 
$\Phi(\textbf{x})$ given by eq.~\eqref{Phi}. Its ground state and excited state are respectively denoted by $|g\rangle$ and $|e\rangle$ with a corresponding 
energy-level spacing $\omega_0$.
In such case, the Hamiltonian of the whole system, detector plus field, can be written as
\begin{eqnarray}
H=H_0+H_\Phi+H_I.
\end{eqnarray}
Here $H_{0}=\frac{1}{2}\omega_0\sigma_z$ is the detector's Hamiltonian, with $\sigma_z$ being the Pauli matrix in the $z$-component, $H_\Phi$ is the field's Hamiltonian, and $H_I$ denotes the interaction Hamiltonian between the detector and field.
Specifically, the interaction Hamiltonian reads
\begin{eqnarray}
\label{I_H}
H_I=\mu\sigma_x\Phi(\textbf{x}),
\end{eqnarray}
where $\mu$ is the coupling constant which is generally assumed to be small and $\sigma_x=\sigma_++\sigma_-$ denotes the Pauli matrix in the $x$-component, with $\sigma_+$$(\sigma_-)$ being the atomic raising (lowering) operator.

At the beginning, we assume that the field is in the Fock vacuum $|0\rangle$, and the detector is prepared at a state, given by $|\Psi(0)\rangle=\cos\frac{\theta}{2}|g\rangle+\sin\frac{\theta}{2}|e\rangle$. Then the total initial state of the detector and field reads $\rho_\text{tot}(0)=\rho(0)\otimes|0\rangle\langle0|$, with $\rho(0)=|\Psi(0)\rangle\langle\Psi(0)|$. The dynamical evolution of the whole system can be described by the Liouville equation. In the interaction picture, it reads
\begin{eqnarray}\label{EE}
	\frac{\partial{\rho_\text{tot}(\tau)}}{\partial\tau}=-i[H_I(\tau),\rho_\text{tot}(\tau)],
\end{eqnarray}
where $\tau$ denotes the proper time of detector. Furthermore, due to the weak coupling assumption, i.e., the coupling strength $\mu$ in Eq.~\eqref{I_H} is small, the Born approximation and the Markov approximation can be applied~\cite{Breuer2002}. In such case,
by tracing over the degrees of freedom of the field we can derive the master equation of the detector that solely describes the detector's dynamics.
The obtained master equation can be written in the Lindblad form:
\begin{eqnarray}\label{LO}
	\nonumber
	\frac{\partial\rho(\tau)}{\partial\tau} &=&-i[H_{\text{eff}},\rho(\tau)]\nonumber\\
    &&+\sum_{j=1}^{3} (2L_j\rho L_j^\dagger-L_j^\dagger L_j\rho-\rho L_j^\dagger L_j),
\end{eqnarray}
where the effective Hamiltonian $H_{\text{eff}}=H_0+H_{\text{shift}}$ includes the Lamb shift \cite{Willis E. Lamb}, $H_\text{shift}=\frac{1}{2}\mu^2\mathrm{Im}(\Gamma_++\Gamma_-)\sigma_z$. The functions $\Gamma_\pm$ will be defined in the following. Moreover, the Lamb shift is typically much smaller than the atomic energy-level spacing $\omega_0$, and thus can be neglected. Recall that the operators in Eq.~(\ref{LO}) are defined as
\begin{eqnarray}
	L_1=\sqrt{\frac{\gamma_+}{2}}\sigma _+,~~~~L_2=\sqrt{\frac{\gamma_-}{2}}\sigma _-,~~~L_3=\sqrt{\frac{\gamma_z}{2}}\sigma_z,
\end{eqnarray}
where
\begin{eqnarray}\label{gamma}
	\nonumber
	\gamma_\pm&=&2\mu^2\mathrm{Re}\Gamma_\pm=\mu ^2\int_{-\infty }^{+\infty} e^{\mp i\omega_0\Delta\tau}G^+(\Delta\tau)d\Delta\tau,
	\\
	\gamma_z&=&0,
\end{eqnarray}
with $\Delta\tau=\tau-\tau^\prime$.
Note that the field Wightman function $G^+(\Delta\tau)$ reads~\cite{V. Husain}
\begin{eqnarray}\label{Wightman function}
G^+(\Delta\tau)&=&\langle 0|\Phi(t(\tau), \textbf{x}(\tau))\Phi(t^\prime(\tau^\prime), \textbf{x}^\prime(\tau^\prime))|0 \rangle\nonumber\\
&=&\int d^3 \mathbf{k} \, |\rho_{|\mathbf{k}|}|^2 e^{i \mathbf{k} \cdot (\mathbf{x} - \mathbf{x}') - i \omega_{|\mathbf{k}|} (t - t' - i \epsilon)},\nonumber\\
\end{eqnarray}
where the distributional character is encoded in the limit $ \epsilon \to 0_+$, and $(t(\tau), \textbf{x}(\tau))$ denotes the worldline of the detector.
Actually, $\gamma_\pm$ are the Fourier transform of the Wightman function, representing the detector's transition rates.

By solving the master equation \eqref{LO} with the initial state $|\Psi(0)\rangle$, one can find that 
the time-dependent reduced density matrix of the detector $\rho(\tau)$ can be written as
\begin{eqnarray}\label{TS}
\rho(\tau)=\frac{1}{2}\bigg(\mathbf{I}+\sum^3_{i=1}\omega_i(\tau)\sigma_i\bigg),
\end{eqnarray}
with 
\begin{eqnarray}\label{ISP}
	\nonumber
	\omega_1(\tau)&=&\text{sin}\theta \text{cos}(\omega_0 \tau)e^{-\frac{1}{2} (\gamma _++\gamma _-)\tau},
\\
	\omega_2(\tau)&=&\text{sin}\theta \text{sin}(\omega_0 \tau)e^{-\frac{1}{2} (\gamma _++\gamma _-)\tau},
\\
	\nonumber
	\omega_3(\tau)&=&-\text{cos}\theta e^{-(\gamma _++\gamma _-)\tau}+\frac{\gamma _+-\gamma _-}{\gamma _++\gamma _-}[1-e^{-(\gamma _++\gamma _-)\tau}].
\end{eqnarray}
Note that the field properties are encoded in $\gamma_\pm$. Therefore, these will affect the evolution of the detector, and can be read out from the measurement of the detector finally.

\section{Quantum coherence dynamics in the Lorentz-invariant and Lorentz-violating field cases}
\label{section3}
In this section, we investigate how the Lorentz violation affects on the behavior of quantum coherence of the UDW detector. Specifically, 
we will consider an inertial detector that moves with a constant velocity in a quantum field vacuum. Its corresponding trajectory reads
\begin{eqnarray}\label{WL}
	(t(\tau),\textbf{x}(\tau))=(\tau \text{cosh}\beta,0,0,\tau \text{sinh}\beta),
\end{eqnarray}
where $\beta$ is the rapidity with respect to the distinguished inertial frame. Substituting this trajectory in Eq.~\eqref{WL} into Eq.~\eqref{gamma}, one can obtain the transition rates of the detector as:
\begin{eqnarray}\label{RF}
	\gamma_\pm&=&
    \frac{\mu^2M_\star}{2\pi\text{sinh}\beta}\int_{0}^{\infty}dg|d(g)|^2 \nonumber
\\
&& \times\mathrm{H}(g\sinh\beta-|(\pm\omega_{0}/M_{*})+gf(g)\cosh\beta|),\nonumber
\\
\end{eqnarray}
where $\mathrm{H}(x)$ is the Heaviside function. For convenience in the following, we define $\gamma_{\pm}=\mu^{2}M_{*}F_{\pm}=\gamma_0c_\pm$ with
\begin{eqnarray}
F_{\pm} &=& \frac{1}{2\pi\sinh\beta}\int_{0}^{\infty}dg|d(g)|^{2} \nonumber
\\
&&\times\mathrm{H}(g\sinh\beta-|h+gf(g)\cosh\beta|).
\end{eqnarray}
Here, the dimensionless argument $h=\pm\omega_0/M_\star$ with the scale parameter $M_*$ entering Eq.~\eqref{RF} only through the overall factor and the ratio $\omega_0/M_\star$. Note that $``+"$ and $``-"$ denotes the excitation and deexcitation rates, respectively.
Furthermore, the dimensionless functions $c_\pm=\pm\frac{2\pi F_\pm}{h}$ are dominated by the rapidity $\beta$ and the dimensionless argument $h$ of the detector, and $\gamma_0={\mu^2\omega_0}/{2\pi}$ denotes the spontaneous emission rate of the detector coupled to the usual massless scalar field. With these parameters, we will discuss how the UDW detector moving with constant velocity responds to the Lorentz-violating quantum field theory in what follows.

\subsection{The Lorentz-invariant scenario}
First, let us consider the Lorentz-invariant field case, where $f(g)=1$ in Eq.~\eqref{RF} has to be imposed. We can find for this case $c_+=0$ and $c_-=1$, i.e., $\gamma_+=0$ and $\gamma_-=\gamma_0$, indicating that the detector cannot be excited while its deexcitation rate remains constant. This result corresponds to the case for Lorentz-invariant field theory. In other words, for the uniformly moving detector that is coupled to the quantum field with Lorentz-invariant dispersion, its transition rates are independent of the detector’s velocity. In this case, according to the definition of quantum coherence measured by $l_1$ norm in Eq.~\eqref{l1}, we can  obtain the corresponding quantum coherence of the uniformly moving detector as 
\begin{eqnarray}
	\label{QC0}
C_{l_1}^0=|\rm{sin}\theta|e^{-\frac{1}{2}\gamma_0\tau},
\end{eqnarray}
where the superscript ``0" indicates the case for the usual quantum field. It is found that the quantum coherence for the uniformly moving detector that is coupled to a Lorentz-invariant quantum field is independent of the detector's rapidity.

\begin{figure}
\includegraphics[width=0.39\textwidth]{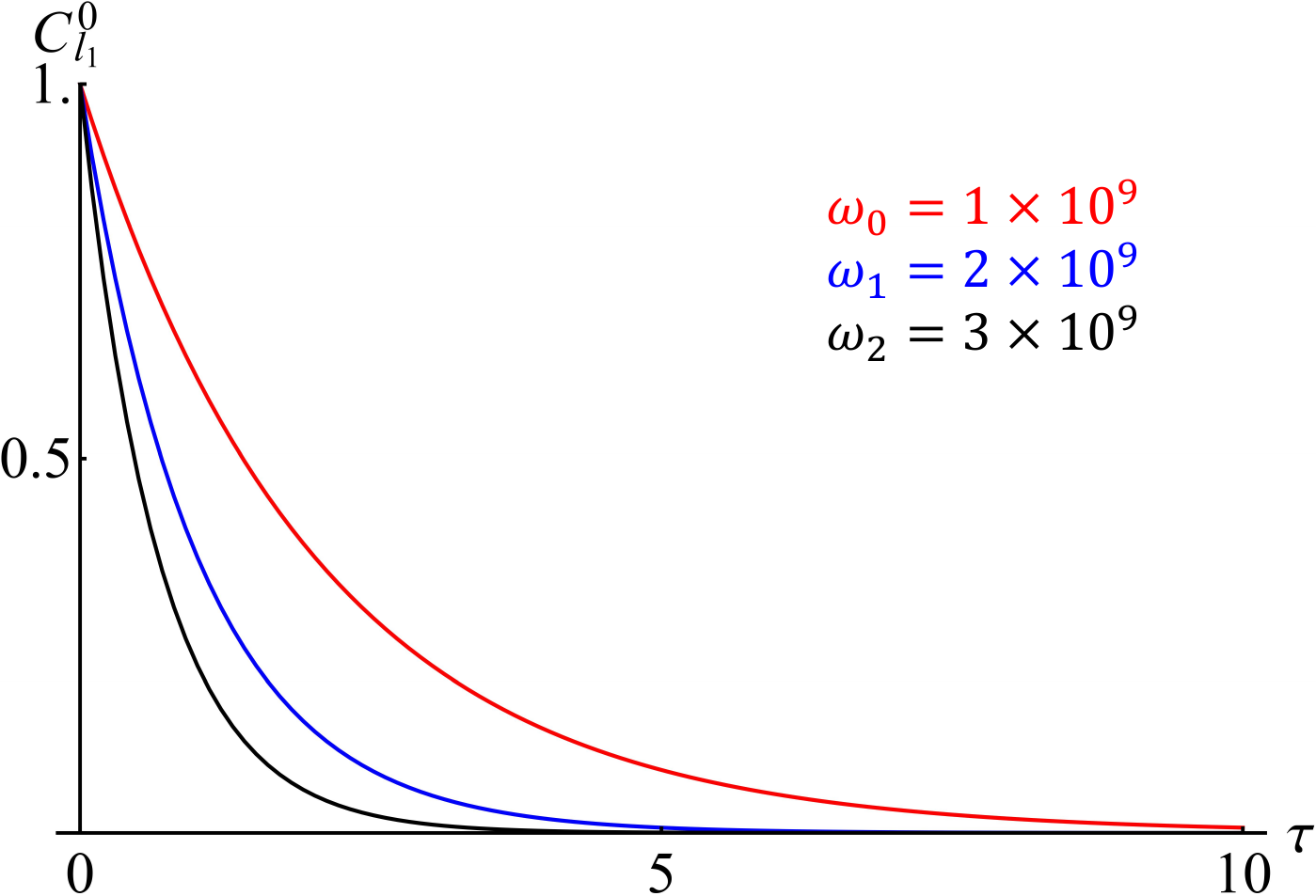}
\caption{Quantum coherence $C_{l_1}^0$ of the UDW detector coupled to the usual massless quantum field. Here, \rm{energy-level spacing} $\omega_0=(10^9, 2\times10^9, 3\times10^9)\rm{Hz}$ correspond to red, blue and black curves, respectively, the initial state $\theta=\pi/2$ and the coupling strength $\mu=10^{-3}$ have been taken.}\label{fig1}
\end{figure}
The corresponding behavior of the quantum coherence $C_{l_1}^0$ for the Lorentz-invariant case is shown in Fig.~\ref{fig1}, which is plotted as a function of the  evolution time $\tau$ (Note that here a scalar transformation of time $\gamma_0\tau$ has been done, while we still denote it as $\tau$ for convenience) with different detector energy-level spacing. The red, blue and black curves correspond to the energy-level spacing $\omega_0=(10^9, 2\times10^9, 3\times10^9) \rm{Hz}$, respectively. Here, we choose the initial state $\theta=\pi/2$ and the coupling strength $\mu=10^{-3}$. It is shown that a larger energy-level spacing leads to the faster decay of quantum coherence. It is worth noting that the $l_1$ norm of quantum coherence is independent of the detector's moving velocity for the Lorentz-invariant case, which is quite different from the Lorentz-violating case that we will analyze below.

\subsection{The Lorentz-violating scenario}
We now consider a uniformly moving UDW detector coupled to a quantum field with the Lorentz-violating dispersion given by Eq.~\eqref{omega}. In this case, according the measurement of quantum coherence in Eq.~\eqref{l1}, we can obtain the corresponding quantum coherence of the detector as
\begin{eqnarray}
\label{QC2}
C_{l_1}^{\mathrm{LIV}}=|\rm{sin}\theta|e^{-\frac{1}{2}\gamma_0(c_++c_-)\tau},
\end{eqnarray}
where the superscript ``$\mathrm{LIV}$" indicates the Lorentz-violating field theory.
The analytical form of the transition rates $\gamma_\pm$ in Eq.~\eqref{RF} features the Heaviside function $\rm{H}(x)$ which is a function of rapidity $\beta$. Therefore, we can find that the behavior of functions $c_\pm$ is inherently rapidity-dependent for the Lorentz-violating case. This dependence is consequently embedded in the quantum coherence described by Eq.~\eqref{QC2}, which is a consequence of the Lorentz violation in quantum field theory.

Our analysis thus reveals a distinction in the response of quantum coherence to detector’s rapidity between the Lorentz-invariant and Lorentz-violating scenarios:
in the former case, the quantum coherence is rapidity-independent, while in the latter case, it depends on the detector’s rapidity. This different response results from
the distinct deexcitation/excitation rates of detector when it is coupled to different quantum fields. In the Lorentz-invariant quantum field theory, the UDW detector's deexcitation/excitation rates are Lorentz-invariant, while in the Lorentz-violating quantum field theory, they are not and show the rapidity-dependence features.
Therefore, whether quantum coherence of UDW detector exhibits rapidity-dependence can serve as a criterion to characterize Lorentz violation in quantum field theory.
Quantum coherence has been extensively studied and well-established measurement techniques are now commonly available \cite{Streltsov2017}.
Therefore, here the rapidity-dependent feature in principle provides a promising observable for detecting the Lorentz violation under current experimental conditions. 
To illustrate the practical operation of this criterion and observable, we next apply the above analysis to a concrete example: the detection of polymer quantization, a framework motivated by loop quantum gravity.

\section{Quantum coherence in polymer quantum field theory} \label{section4}
In this section, we will explore the polymer quantization method that arises in loop quantum gravity~\cite{Rovelli2007,Thiemann2008}, which is a background-independent canonical quantization of general relativity and a well-recognized candidate theory for quantum gravity~\cite{Ashtekar2004}.
Beyond its origins in quantum gravity, this polymer quantization scheme has also been applied to a lot of physical systems, including the mechanical systems and the scalar field~\cite{1Ashtekar2003,2Ashtekar2003,1Husain2010,2Husain2010,3Husain2010,Seahra2012}. There are two main defining features that distinguish the polymer quantization from the usual Schr\"{o}dinger quantization: (i) In addition to the Planck constant $\hbar$, it introduces a new dimensional parameter, i.e., the Planck length $L_p=\sqrt{\hbar G/c^3}$ (where $G$ denotes Newton’s gravitational constant, and $c$ is the speed of light in vacuum) in the context of quantum gravity;
(ii) Position and momentum operators cannot be simultaneously well-defined in polymer quantization, arising from the nonseparable nature of the kinematical Hilbert space. Given these differences, it is natural to anticipate that a polymer quantization will lead to different results compared to the usual Schr\"{o}dinger quantization.

We adopt the specific implementation of a polymer quantized scalar field studied in Ref.~\cite{2Husain2010}, referring therein for details. This polymer quantization framework has been extensively studied and fruitfully applied in exploration of the transition rates of UDW detector, including 
the inertial detector cases~\cite{1Husain2016,Kajuri2016,Kajuri2018,Louko2018} and accelerated detector cases~\cite{Husain2015,Stargen2017}.
In the present work, we will investigate the dynamics of quantum coherence of UDW detector that is coupled to a polymer quantized scalar field.
We mainly focus on the influence of polymer quantization on the quantum coherence, as well as the responses of quantum coherence to the detector's relativistic inertial motion. In particular, we shall establish both the critical rapidity for sharped response of UDW detector's quantum coherence to the polymer quantized scalar field theory
and the magnitude of the change.

\subsection{Transition rates in the polymer quantized field theory}\label{Transition rates}
As shown in Ref.~\cite{2Husain2010}, the Wightman function in polymer quantized field theory is given by
\begin{eqnarray}\label{WF}
	\nonumber
	G^+(t, \textbf{x}; t^\prime, \textbf{x}^\prime)&=&\frac{1}{(2\pi)^3} \int d^3\textbf{k}e^{i\textbf{k}\cdot({\textbf{x}-\textbf{x}^\prime)}}
	\\
	&&\times \sum_{n=0}^{\infty } |c_{n}(|\textbf{k}|)|^2e^{-i\Delta E_{n}(|\textbf{k}|)(t-t^\prime)},
\end{eqnarray}
where $\Delta E_n(|\bk|)=E_n(|\bk|)-E_0(|\bk|)$ with
\begin{eqnarray}
	E_{2n}(|\textbf{k}|)/\omega&=&\frac{2g^2A_n(1/4g^2)+1}{4g},
	\\
	E_{2n+1}(|\textbf{k}|)/\omega&=&\frac{2g^2B_{n+1}(1/4g^2)+1}{4g},
	\\
	c_{n}(|\textbf{k}|)&=&1/\sqrt{M_\star}\int_{0}^{2\pi}  \psi_n(i\partial_{u})\psi_0du.
\end{eqnarray}
Furthermore, $\psi_n$ in the above formula are given by 
\begin{eqnarray}
	\psi_{2n}(u)=\pi^{-1/2}\mathrm{ce}_{n}(1/4g^2,u),
\end{eqnarray}
\begin{eqnarray}
	\psi_{2n+1}(u)=\pi^{-1/2}\mathrm{se}_{n+1}(1/4g^2,u),
\end{eqnarray}
which satisfies the Matheieu equation with $\mathrm{ce}_{n}(x,q)$ and $\mathrm{se}_{n}(x,q)$ being the elliptic cosine and sine functions, respectively, and $B_{n}(x)$ and $A_{n}(x)$ are the Mathieu characteristic value functions.

In order to learn the dynamics of the UDW detector moving inertially in the polymer quantized field theory, we have to know the detector's transition rates in this scenario.
Substituting the inertial trajectory of UDW detector in Eq.~\eqref{WL} into the Wightman function in Eq.~\eqref{WF}, we can straightly calculate the corresponding transition rates in Eq.~\eqref{gamma},

\begin{eqnarray}\label{TRRs1}
	\nonumber	\gamma_\pm^{\mathrm{PQ}}&=&\frac{\mu^2M_\star}{2\pi\text{sinh}\beta}\sum_{n=0}^{\infty}\int_{0}^{\infty}dgg|\sqrt{M_\star}c_{n}(g)|^2
	\\
	&&\times \text{H}(g\sinh\beta-| h+\omega_n(g)\cosh\beta|),
\end{eqnarray}
where $\omega_{n}(g)=\Delta E_{n}(g)/M_\star$, and the superscript ``$\mathrm{PQ}$" indicates the polymer quantized scalar
field theory. As considered in Ref.~\cite{2Husain2010}, only the matrix elements $c_{4n+3}$ (for $n=0, 1, 2, \cdots$) are nonvanishing. This key property simplifies the calculation of transition rates.
By substituting this nonvanishing condition into Eq.~\eqref{TRRs1}, we can rewrite the transition rates as $\gamma_\pm^{\mathrm{PQ}}=\mu^{2}M_{*}F^\text{PQ}_{\pm}$ with
\begin{eqnarray}\label{TRRs2}
	F_\pm^{\mathrm{PQ}}&=&\frac{1}{2\pi\text{sinh}\beta}\sum_{n=0}^{\infty}\int_{0}^{\infty}dgg|\sqrt{M_\star}c_{4n+3}(g)|^2\nonumber
	\\
	&&\times \text{H}(g\sinh\beta-| h+\omega_{4n+3}(g)\cosh\beta|).
\end{eqnarray}

\begin{figure}
\centering
\includegraphics[width=0.39\textwidth]{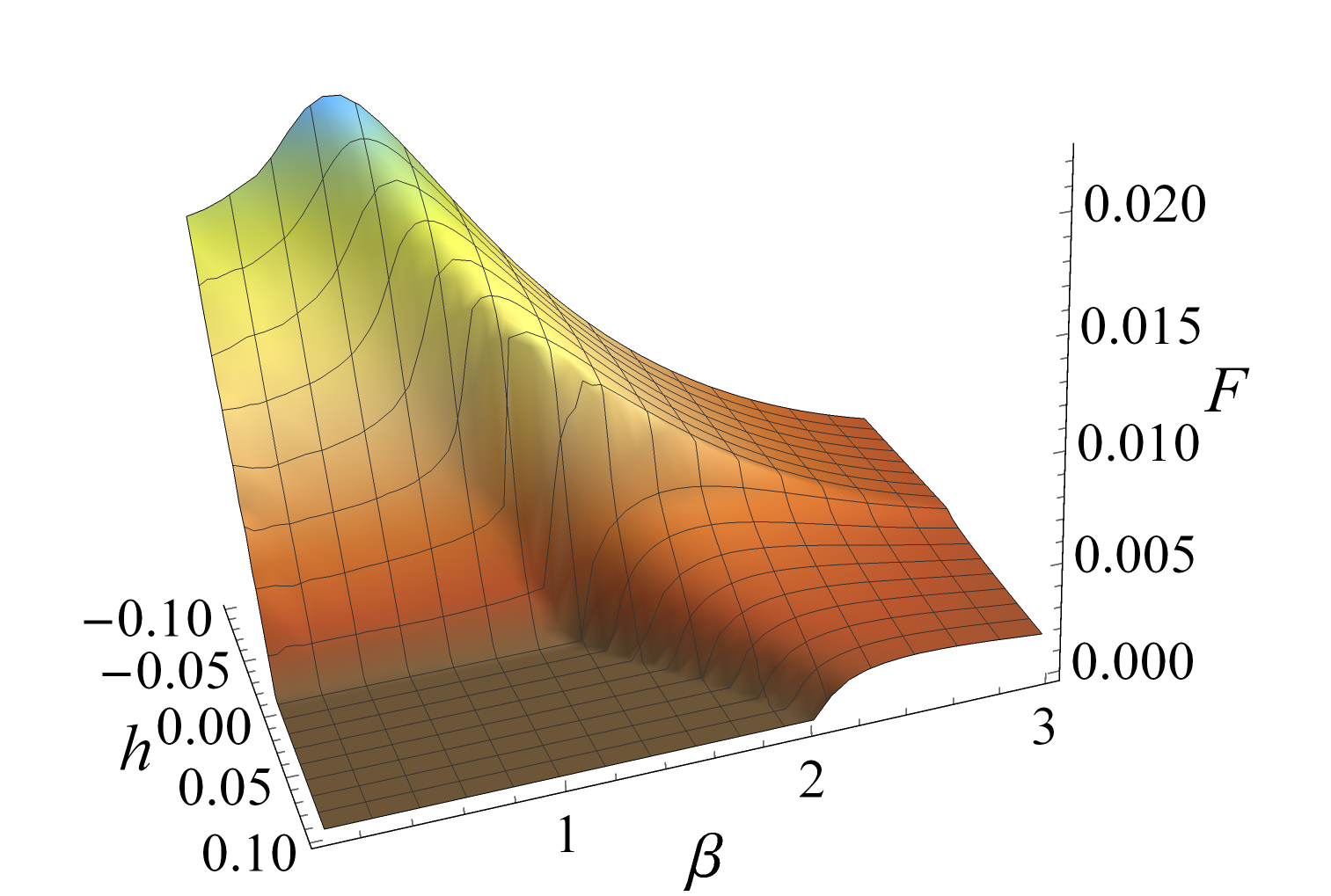}
\caption{Transition rates $F_\pm^{\mathrm{PQ}}$ in the polymer quantized field theory as functions of detector's rapidity $\beta$  and energy-level spacing $h$. The positive and negative parts at $h$ axis correspond to excitation and deexcitation rates, respectively. }\label{fig2}
\end{figure}

Based on the previous analysis, we can find that when a detector is coupled to a usual massless quantum scalar field, its transition rates are independent of the detector's rapidity $\beta$. However, here we can find that the transition rates in Eq.~\eqref{TRRs2} for the polymer quantized field theory case depend not only on $\beta$ but also on the energy-level spacing $h$. To better understand the behavior of the transition rates in polymer quantized field theory, we plot the transition rates $F_\pm^\mathrm{PQ}$ as functions of both the detector's rapidity $\beta$ and the energy-level spacing $h$ in Fig.~\ref{fig2}. 
Note that the positive and negative parts on the $h$ axis correspond to the excitation and deexcitation rates, respectively. From this plot, we can find that there exists a critical value of rapidity: below this value, the excitation rate vanishes, while beyond it, the detector becomes excited automatically.
This characteristic can be attributed to the dispersion relation of the polymer field, where the function $f$ can drop below unity in certain regions~\cite{2Husain2010}. The numerical calculation shows that the critical value is $\beta_c\approx1.3675$. Moreover, the detector deexcitation rate never vanishes, while is still affected by rapidity $\beta$.

\subsection{Quantum coherence in the polymer quantized quantum field theory}
According to the transition rates in Eq.~\eqref{TRRs2}, the corresponding dimensionless functions $c_\pm^{\mathrm{PQ}}=\pm\frac{2\pi}{h}F_\pm^{\mathrm {PQ}}$ can be written as 
\begin{eqnarray}\label{RF2}
\nonumber	
c_\pm^{\mathrm{PQ}}&=&
\frac{\pm1}{h\;\text{sinh}\beta}\sum_{n=0}^{\infty}\int_{0}^{\infty}dgg|\sqrt{M_\star}c_{4n+3}(g)|^2
	\\
	&&\times \text{H}(g\sinh\beta-|h+\omega_{4n+3}(g)\cosh\beta|).
\end{eqnarray}
Then, applying Eq. \eqref{RF2} to the general expression for the $l_1$ norm of quantum coherence in Eq. \eqref{l1}, we can derive the explicit form of $C_{l_1}^{\mathrm{PQ}}$ in the polymer quantized scalar field theory,
\begin{eqnarray}
\label{QC3}
C_{l_1}^{\mathrm{PQ}}=|\rm{sin}\theta|e^{-\frac{1}{2}(c_+^{\mathrm{PQ}}+c_-^{\mathrm{PQ}})\gamma_0\tau}.
\end{eqnarray}
It should be noted that the maximum quantum coherence is achieved when $\theta=\pi/2$. Furthermore, the exponential term indictates the temporal evolution of quantum coherence with time, and its form directly reflects the influence of the polymer quantization effects. These effects are fully encoded in the dimensionless  functions $c_\pm^{\mathrm{PQ}}$ given by Eq.~\eqref{RF2}.

\begin{figure*}
\includegraphics[width=0.68\textwidth]{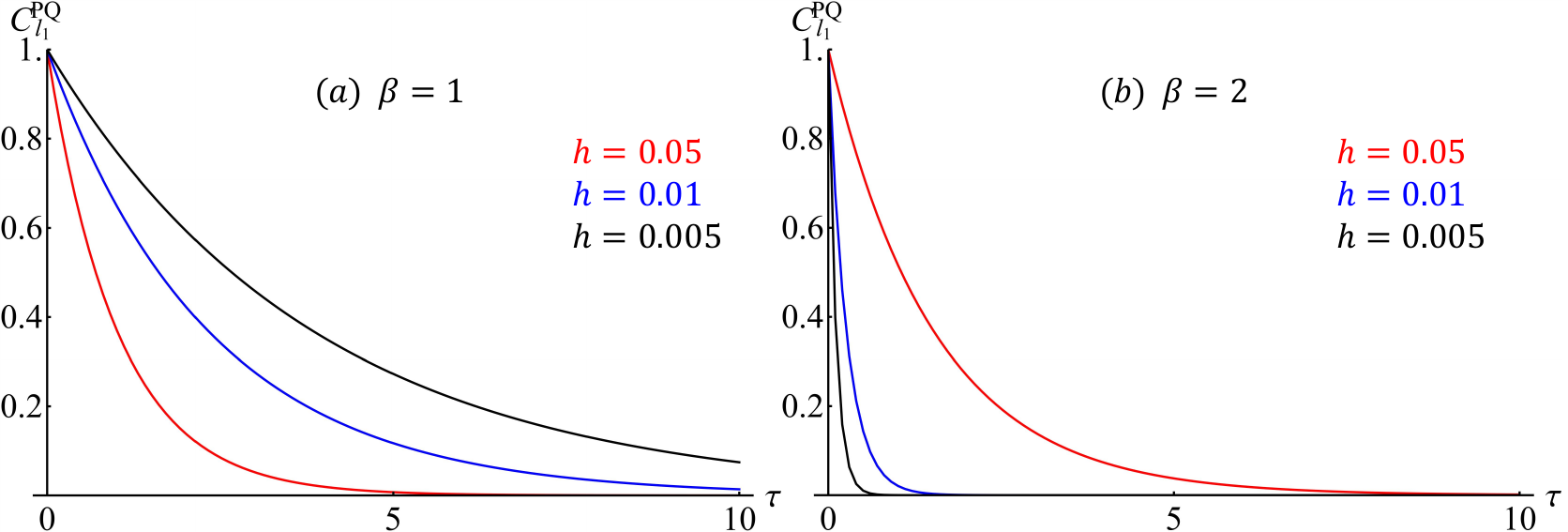}
\caption{Time evolution of quantum coherence $C_{l_1}^{\mathrm{PQ}}$ with different $h$ for the rapidity $\beta$ below (left) and above (right) the critical value $\beta_c$ cases. Here $h=(0.05, 0.01, 0.005)$ corresponding to red, blue and black curve, respectively. Energy-level spacing $\omega_0=10^9\rm{Hz}$, initial state $\theta=\pi/2$ and
coupling strength $\mu=10^{-3}$ have been taken.}\label{fig3}
\end{figure*}

\begin{figure*}
\includegraphics[width=0.92\textwidth]{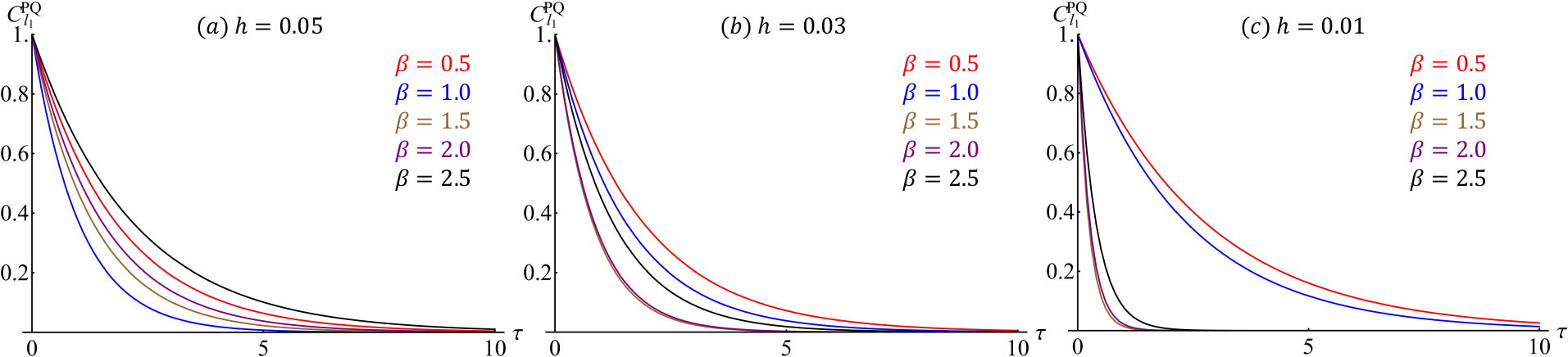}
\caption{Time evolution of quantum coherence $C_{l_1}^{\mathrm{PQ}}$ for several
values of rapidity $\beta$ with different $h$. Here $\beta=(0.5, 1.0, 1.5, 2.0, 2.5)$ corresponding to red, blue, brown, purple and black curve, respectively. Energy-level spacing $\omega_0=10^9\rm{Hz}$, initial state $\theta=\pi/2$ and
coupling strength $\mu=10^{-3}$ have been taken. }\label{fig4}
\end{figure*}

In what follows, will we investigate how two key physical parameters, i.e., the detector's rapidity $\beta$ and the effective polymer energy scale ratio $h$, influence the quantum coherence of the UDW detector interacting with a polymer quantized quantum field.
As discussed in Sec.~\ref{Transition rates}, the transition rates exhibit qualitatively different behaviors below and above the critical rapidity $\beta_c$ for arbitrarily small $h$. To make this dependence explicitly, in Fig.~\ref{fig3} we present the time evolution of the $l_1$ norm of quantum coherence $C_{l_1}^{\mathrm{PQ}}$ for different $h$ values in two regimes: $\beta < \beta_c$ (left panel) and $\beta > \beta_c$ (right panel). 
The parameters are fixed as $\omega_0 = 10^9\mathrm{Hz}$, $\theta = \pi/2$, and $\mu = 10^{-3}$. 
From Fig.~\ref{fig3}, we can find that for $\beta < \beta_c$ case, smaller $h$ leads to a slower decay of coherence. 
This behavior arises because the excitation process is suppressed ($c_+^{\mathrm{PQ}} = 0$), and only the deexcitation rate $c_-^{\mathrm{PQ}}$ contributes to the coherence dynamics. Since $c_-^{\mathrm{PQ}}$ increases with $h$, a larger $h$ enhances dissipation and thus accelerates the loss of coherence.
In contrast, for $\beta > \beta_c$ case, the situation reverses: smaller $h$ results in faster decoherence. This change is attributed to the appearance of nonvanishing excitation rates ($c_+^{\mathrm{PQ}} \neq 0$) above $\beta_c$, where both $c_+^{\mathrm{PQ}}$ and $c_-^{\mathrm{PQ}}$ increase as $h$ decreases. The simultaneous enhancement of excitation and deexcitation processes thus amplifies dissipation, leading to a more rapid decay of quantum coherence at smaller $h$.

To explore how the motion of detector influences quantum coherence, in Fig.~\ref{fig4} we plot the quantum coherence $C_{l_1}^{\mathrm{PQ}}$ as a function of the evolution time for different rapidity $\beta$ by fixing $h$. We can find that for fixed $h$, the $l_1$ norm decays more rapidly as the rapidity increases when $\beta < \beta_c$. This is because that in this regime, the inertial detector cannot be excited, i.e., $c_+^{\mathrm{PQ}} = 0$ in such case, while only the deexcitation rate $c_-^{\mathrm{PQ}}$ for the detector exists  and increases with the detector's rapidity $\beta$. 
Therefore, higher rapidity $\beta<\beta_c$ can enhance dissipation and accelerate the loss of coherence.
Once the rapidity exceeds the critical value $\beta_c$, we can see from Eq. \eqref{TRRs1} that it is possible for the detector to be excited, i.e., $c_+^{\mathrm{PQ}} \neq 0$, under certain condition. In such case, both the excitation rate, $c_+^{\mathrm{PQ}}$, the deexcitation rate $c_-^{\mathrm{PQ}}$ exist and 
have a combined effect on the dynamics of quantum coherence. 
Interestingly, we can find that when $\beta$ exceeds a certain value beyond $\beta_c$, the quantum coherence decreases more slowly with the increase of 
the rapidity. This means in this rapidity regime (exceeding a certain value beyond $\beta_c$) the competition between excitation and deexcitation processes 
can suppress decoherence to some extent. It is also worth noting that the sensitivity of $C^{\mathrm{PQ}}_{l_1}$ to rapidity becomes more pronounced as $h$ decreases. 
Compared with the case of larger $h$ (left panel), smaller $h$ (right panel) leads to sharper variations of $C^{\mathrm{PQ}}_{l_1}$ near the transition region around 
$\beta_c$. This tendency is consistent with our earlier analysis: near the critical rapidity, the effective transition rates $c^{\mathrm{PQ}}_\pm$ exhibit more distinct variations with respect to $h$ on either side of $\beta_c$, which amplifies the change of coherence across the critical point for smaller $h$.

\begin{figure}
\includegraphics[width=0.38\textwidth]{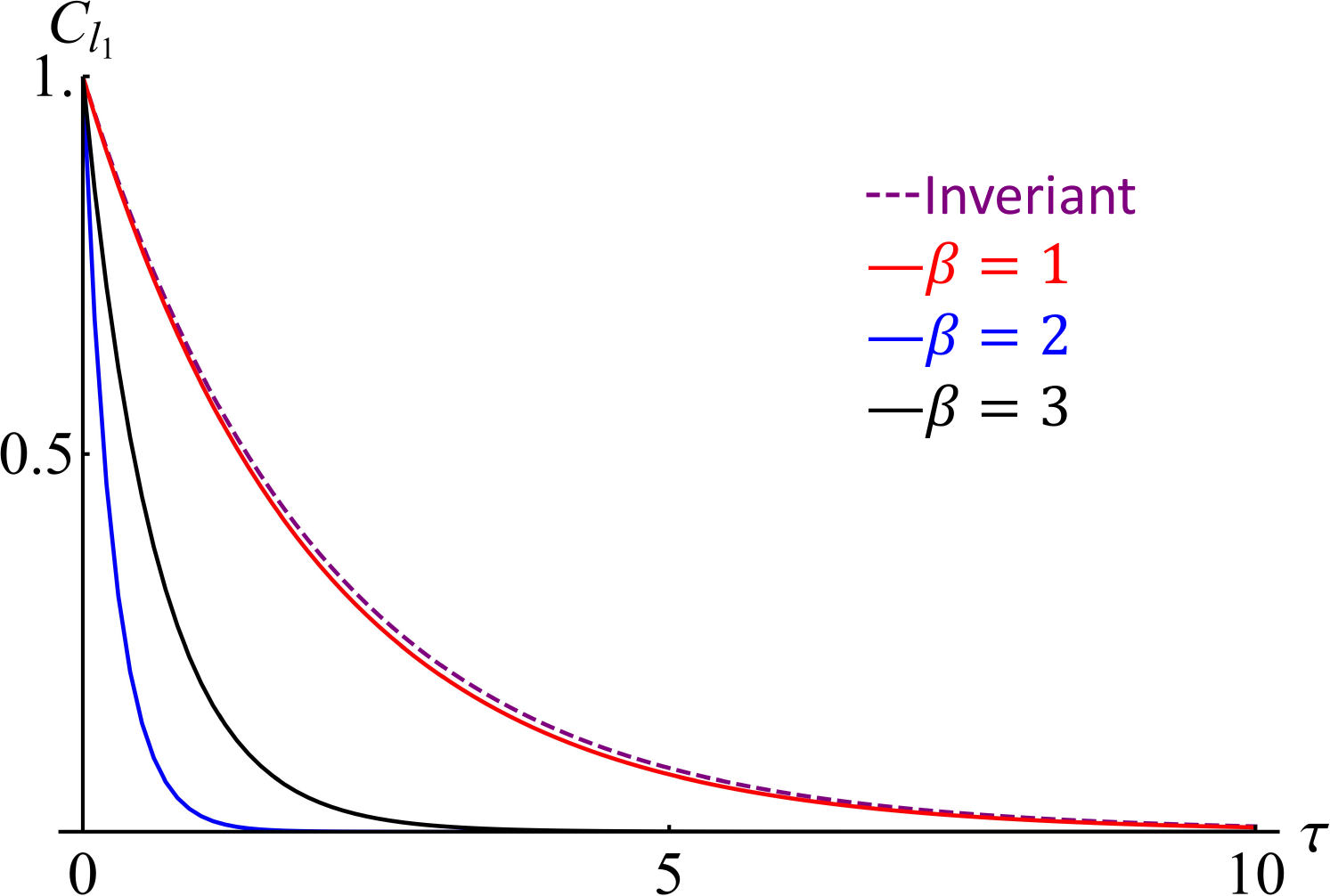}
\caption{$l_1$ norm in polymer quantized field case as a function of evolution time $\tau$ with different $\beta$ compared with Lorentz-invariance case. Here real lines are $l_1$ norm of polymer quantization case, $\beta=(1, 2, 3)$ corresponding to red, blue and black curve respectively, the dashed purple curve is Lorentz-invariant case. Energy-level spacing $\omega_0=10^9\rm{Hz}$, initial state $\theta=\pi/2$,
coupling strength $\mu=10^{-3}$ and $h=0.01$ have been taken. }\label{fig5}
\end{figure}

In order to highlight the unique properties of polymer quantization, we compare the quantum coherence dynamics in the polymer quantized field theory with that in the Lorentz-invariant case. Unlike the Lorentz-invariant scenario, where the quantum coherence is independent of the detector's rapidity, the polymer quantized field theory introduces a clear rapidity-dependence due to its modified dispersion relation rooted in nonperturbative quantum gravity effects. 
Fig.~\ref{fig5} shows the $l_1$ norm $C_{l_1}$ as a function of the  evolution time $\tau$ for different rapidity $\beta=(1,2,3)$, where
the red, blue, and black solid curves represent the polymer quantization case, and the dashed purple curve denotes the Lorentz-invariant case as reference.
It is evident that for the polymer quantized field case, the coherence decay exhibits a nonmonotonic dependence on $\beta$: the $l_1$ norm first decreases and then increases as the detector's rapidity increases. 
This behavior arises from the rapidity-dependent transition rates, where the combined rate $c_+^{\mathrm{PQ}} + c_-^{\mathrm{PQ}}$ initially increases and then decreases with $\beta$. Therefore, by appropriately tuning the detector's rapidity, one can determining which kind of quantum field the detector is coupled to through 
checking whether the behavior of quantum coherence depends on the rapidity.

\begin{figure}
\includegraphics[width=0.38\textwidth]{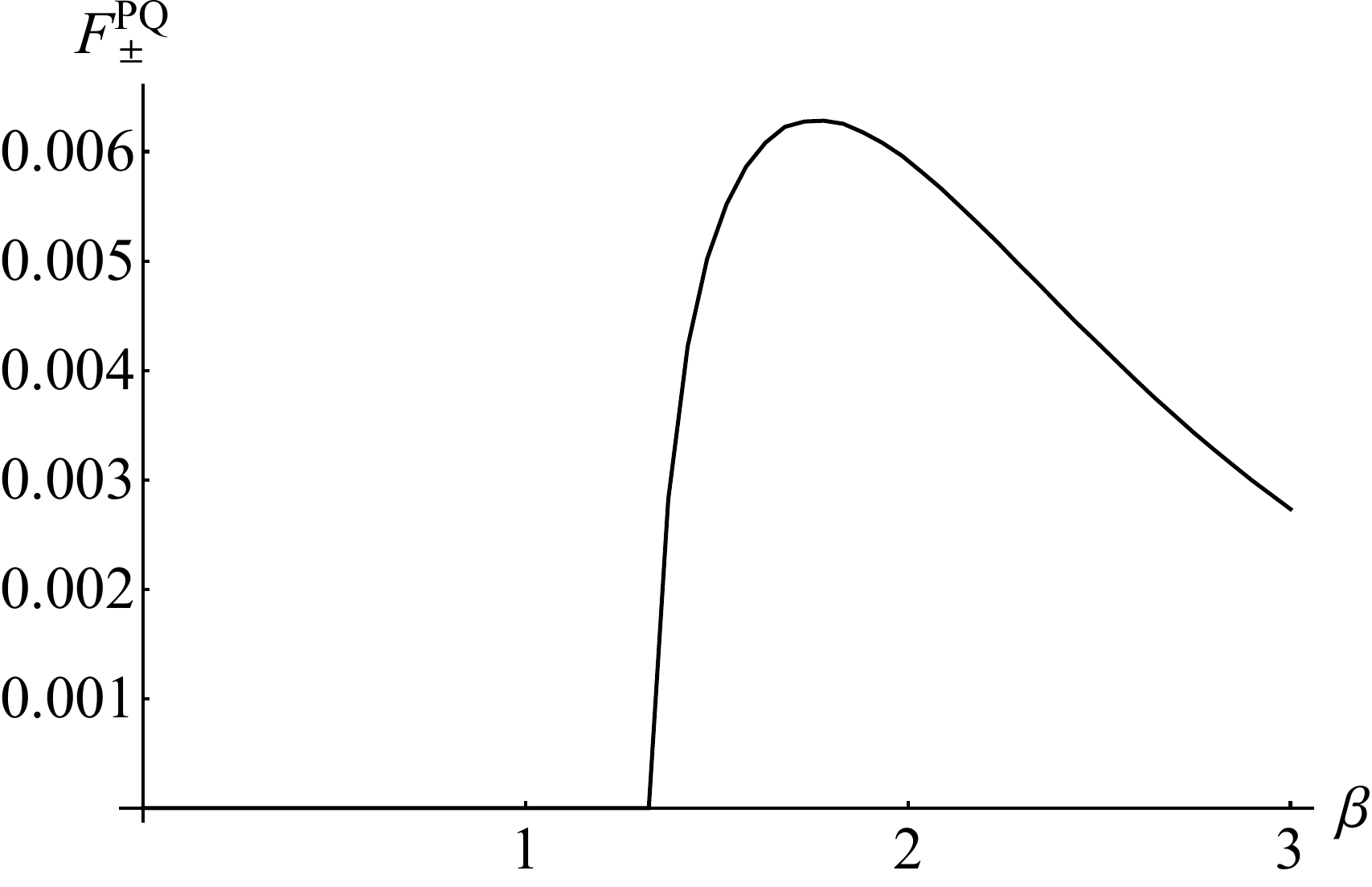}
\caption{The transition rates $F^\text{PQ}_\pm$ as a function of the detector's rapidity in the limit $h\rightarrow0$. }\label{fig6}
\end{figure}

Our UDW detector model is quite similar to the dipole moment interaction by which an atom couples the quantized electromagnetic field \cite{Eduardo2013}. 
Therefore, our study should apply to atoms or ions moving with a relativistic velocity, including the ions accelerated to rapidity $\beta\approx3$ at the RHIC.
We will discuss the relevant implementation, through employing the rapidity-dependent coherence signature to probe the polymer-quantized field under realistic conditions. In Ref.~\cite{A. Garcia-Chung}, the polymer energy scale is constrained within the range $M_\star \in [10^{13}, 10^{25}]~\mathrm{GeV}$, implying that the dimensionless parameter $h$ is extremely small, typically $h \in [10^{-40}, 10^{-28}]$. In this limit, the excitation and deexcitation rates become nearly identical. In Fig.~\ref{fig6} we show these transition rates $F_\pm^{\mathrm{PQ}}$ as a function of detector's rapidity $\beta$ in the limiting $h \to 0$. Note that the actual transition rates asymptotically approach this curve rather than coincide with it, since $h$ can be very small but never exactly zero.
Two key features can be identified. (i) When $\beta < \beta_c$, both $F_+^{\mathrm{PQ}}$ and $F_-^{\mathrm{PQ}}$ approach zero, leading to negligible transition rates. (ii) Near $\beta_c$, both functions exhibit a sharp peak, corresponding to a sudden enhancement in the detector's transition probabilities.
 Although $F_\pm^{\mathrm{PQ}}$ remains finite across $\beta_c$, according to the definition $c_\pm^{\mathrm{PQ}} = 2\pi F_\pm^{\mathrm{PQ}} / h$, the small $h$ amplifies $c_\pm^{\mathrm{PQ}}$ dramatically, generating a Lorentz-like burst in the transition rates.
Consequently, the quantum coherence, which depends exponentially on $c_\pm^{\mathrm{PQ}}$, displays a pronounced discontinuity near the critical rapidity in the polymer quantization case.
Importantly, the critical rapidity $\beta_c \simeq 1.3675$ lies within the accessible range of current accelerator facilities, such as the Relativistic Heavy Ion Collider~\cite{velocity}.
This suggests that, in principle, Lorentz violation effects induced by polymer quantization could be experimentally detectable through coherence measurements near $\beta_c$. In the following analysis, we take $M_\star = 10^{11} \mathrm{GeV}$ as a representative example to illustrate this behavior explicitly, noting that the true polymer scale is expected to be considerably higher.

\begin{figure}
\includegraphics[width=0.42\textwidth]{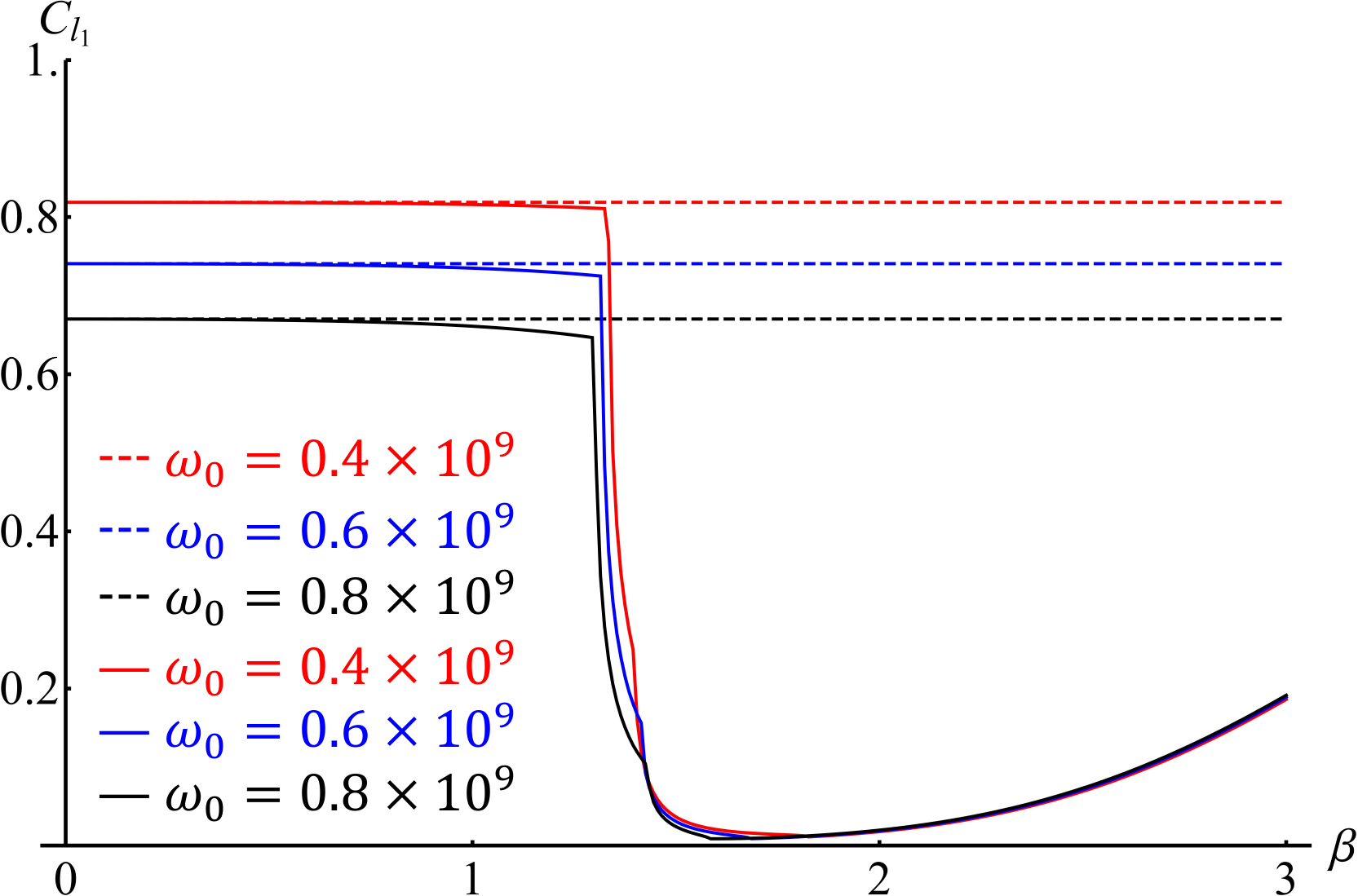}
\caption{$l_1$ norm  coherence as a function of detector's rapidity $\beta$ with different energy-level spacing $\omega_0$. The solid curves denote the $l_1$ norm coherence for the polymer quantized field case, dashed curves are the $l_1$ norm coherence for the Lorentz-invariant case. Here, $\omega_0=(0.4\times10^9, 0.6\times10^9, 0.8\times10^9)\rm{Hz}$ correspond to red, blue and black curve, respectively, the evolution time $\tau=1$, the polymer energy scalar $M_\star=10^{11}\rm{GeV}$ is assumed for an example, the initial state $\theta=\pi/2$ and
coupling strength $\mu=10^{-3}$.}\label{fig7}
\end{figure}

\begin{figure*}
\includegraphics[width=0.36\textwidth]{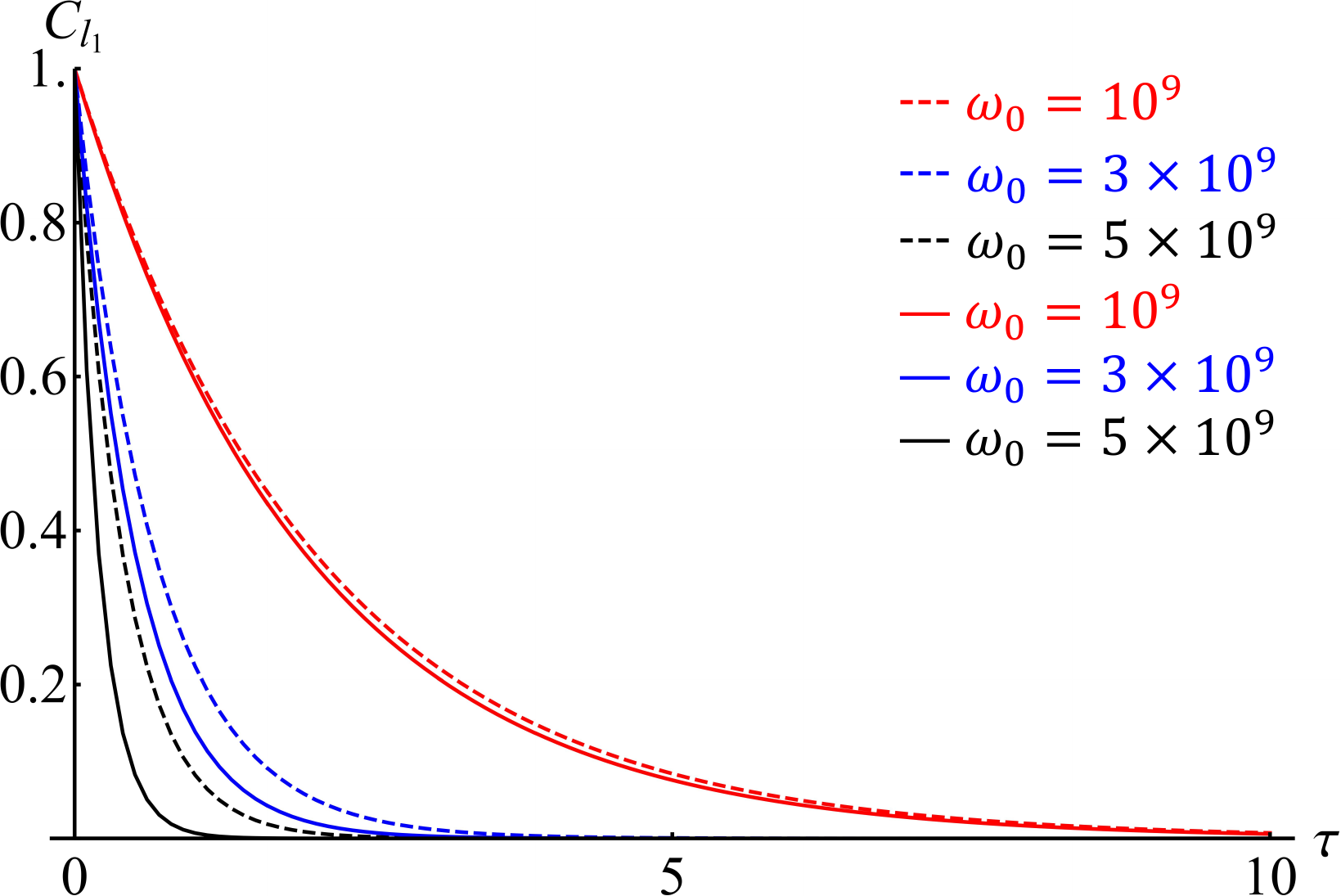}
\includegraphics[width=0.36\textwidth]{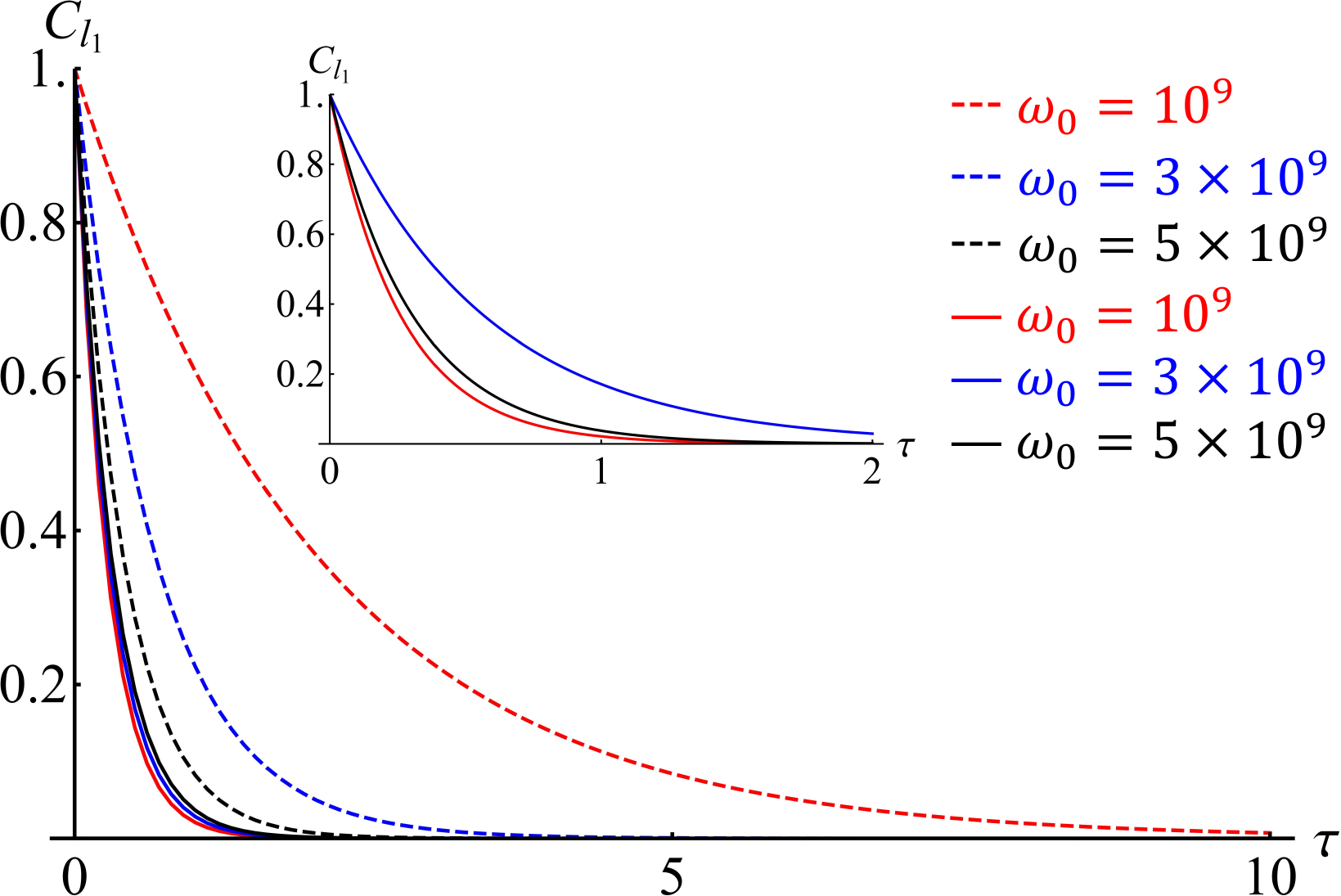}
\caption{$l_1$ norm coherence $C_{l_1}$ as a function of the evolution time $\tau$ with different energy-level spacing $\omega_0$. Here, solid curves are $l_1$ norm coherence for the polymer quantized field case, dashed curves are $l_1$ norm coherence for the Lorentz-invariant field case. $\omega_0=(10^9, 3\times10^9, 5\times10^9)\rm{Hz}$ correspond to red, blue and black curves, respectively. Polymer energy scalar $M_\star=10^{11}\rm{GeV}$ is assumed as an example, rapidity $\beta=1$ (left panel) and $\beta=2$ (right panel), initial state $\theta=\pi/2$ and
coupling strength $\mu=10^{-3}$ have been taken.}\label{fig8}
\end{figure*} 

In order to better understand how the detector's rapidity influences quantum coherence in the polymer quantized field case, and in particular, find its difference 
with that for the Lorentz-invariant scenario, we plot in Fig.~\ref{fig7} the $l_1$ norm of coherence as a function of rapidity $\beta$ with different energy-level spacings $\omega_0$. The solid curves correspond to the polymer quantized field case, while the dashed curves represent the Lorentz-invariant case. In the Lorentz-invariant case, the $l_1$ norm remains independent of $\beta$, consistent with the fact that the transition rates do not depend on the detector's rapidity. In contrast, for the polymer quantized field, the $l_1$ norm remains nearly constant for $\beta < \beta_c$, and exhibits a sudden drop when $\beta$ exceeds the critical rapidity $\beta_c$. This sharp change arises from the rapid variation of the transition rates $c_\pm^{\mathrm{PQ}}$ near $\beta_c$, which originates from the modified field dispersion relation introduced by polymer quantization.
Therefore, a pronounced reduction in the $l_1$ norm coherence serves as a distinctive signature of the polymer quantized field theory. To maximize the detectability of this effect, the detector's rapidity should be tuned as closely as possible to the critical value $\beta_c$, where the deviation from Lorentz-invariant behavior reaches its maximum.

Finally, we will explore how quantum coherence evolves with different energy-level spacing $\omega_0$ for the polymer quantized field case, and compare these results with that for the Lorentz-invariant case. In Fig.~\ref{fig8}, we plot the $l_1$ norm coherence $C_{l_1}$ as a function of evolution time $\tau$ with several values of $\omega_0$. Here, we take two distinct rapidity values: one below $\beta_c$ and the other beyond $\beta_c$.
In the left panel where $\beta=1<\beta_c$,  we can find that the $l_1$ norm coherence exhibits a faster decay as $\omega_0$ increases.
This behavior indicates that a larger energy-level spacing of detector amplifies the decoherence induced by the polymer quantized field, leading to a quicker loss of quantum coherence at fixed rapidity. Consequently, detectors with larger $\omega_0$ should be more sensitive for detecting polymer quantized field theory in the subcritical regime.
In contrast, the right panel shows the supercritical regime $\beta=2>\beta_c$, where the $l_1$ norm coherence displays a nonmonotonic dependence on $\omega_0$, specifically, the strength of decay initially decreases and subsequently increases with the increase of the energy-level spacing.
Besides, we can find that no matter for which cases the quantum coherence for the polymer quantized field case display a more sharp decay than 
that for the Lorentz-invariant field case.
Therefore, to clearly distinguish the polymer field scenario from the Lorentz-invariant case, one must carefully choose a suitable energy-level spacing such that the resulting $l_1$-norm coherence decay deviates most significantly from that in the Lorentz-invariant field theory.

\section{Discussions and Conclusions}
\label{section5}
We have investigated the behavior of quantum coherence for a two-level UDW detector and studied its response to the Lorentz violation in quantum field theory. Specifically, we considered the polymer quantized scalar field as a concrete realization of Lorentz violation and compared its effect on quantum coherence with that of the Lorentz-invariant field case.

For the Lorentz-invariant case, the quantum coherence characterized by the $l_1$ norm remains completely independent of the detector's rapidity, since the corresponding transition rates are rapidity-independent. In contrast, in the polymer quantized field scenario, the quantum coherence exhibits a  rapidity dependence and undergoes a sharp transition near a critical rapidity $\beta_c \simeq 1.3675$. This critical behavior originates from the rapid enhancement of the transition rates around $\beta_c$, which in turn amplifies the rapidity dependence of the dimensionless coefficients $c_\pm$. Because the quantum coherence depends exponentially on $c_\pm$, this amplification results in a sudden change of coherence near the critical value.
Moreover, the detector's energy-level spacing $\omega_0$ plays distinct roles in different rapidity values.
For $\beta < \beta_c$, increasing $\omega_0$ accelerates the decay of quantum coherence due to the monotonic growth of $c_-$ (while $c_+ = 0$ in this case). However, for $\beta > \beta_c$, the decay behavior of quantum coherence exhibits a nonmonotonic dependence on $\omega_0$, first decreases and then increases with the increase of the energy-level spacing $\omega_0$. 
In all cases, the quantum coherence decay induced by polymer quantized field is faster than in the Lorentz-invariant case, and this discrepancy becomes more pronounced for smaller values of the polymer parameter $h$. 
These distinctive quantum coherence behavior could be served as the criteria to determine whether the Lorentz invariance is broken or not.
In particular, the critical rapidity $\beta_c \simeq 1.3675$ lies within the range achievable in current experimental facilities, such as the Relativistic Heavy Ion Collider. Therefore, our results suggest that the predicted quantum coherence behavior could be in principle tested experimentally by controlling the detector's rapidity $\beta$ and tuning energy-level spacing $\omega_0$ appropriately.

This work demonstrates that quantum coherence serves as a highly sensitive probe of Lorentz violation, providing a complementary approach to the tests of Lorentz-invariance violation.
The dependence of quantum coherence signatures on rapidity and energy-level spacing of detector directly encodes the modified dispersion relations of the polymer quantized field, thereby providing a potential window into quantum gravity field theory.
Future extensions may include exploring coherence and entanglement dynamics in multi-detector systems or investigating the dependence on the polymer energy scale $M_\star$, which could further tighten experimental constraints on Lorentz  violation.

\begin{acknowledgments}
ZT was supported by the National Natural Science Foundation of China under Grant No. 12575050, and the scientific research start-up funds of Hangzhou Normal University: 4245C50224204016. This research was supported by Hangzhou Leading Youth Innovation and Entrepreneurship Team project  under Grant No. TD2024005,
and the HZNU scientific research and innovation team project (TD2025003).
XL acknowledges the support by the National Natural Science Foundation of China (NSFC) under Grant No. 12065016 and the Discipline-Team of Liupanshui Normal University of China under Grant No. LPSSY2023XKTD11. 

\end{acknowledgments}

\onecolumngrid
\vspace{1.5cm}

\clearpage
\widetext


\begin{thebibliography}{100}
\bibitem{Mattingly2005} D. Mattingly, Living Rev. Relativity {\bf 8}, 5 (2005).
\bibitem{Amelino2013} G. Amelino-Camelia, Living Rev. Relativity {\bf 16}, 5 (2013).
	
\bibitem{S. Samuel}
V. A. Kosteleck\'{y}, and S. Samuel, Phys. Rev. D {\bf39}, 683 (1989).

\bibitem{Randall1999}
L. Randall and R. Sundrum, Phys. Rev. Lett. {\bf83}, 3370 (1999).
\bibitem{Burgess2002}
C. P. Burgess, Living Rev. Relativity {\bf7}, 5 (2004).

\bibitem{R. Gambini}
R. Gambini, J. Pullin, Phys. Rev. D {\bf59}, 124021 (1999).
\bibitem{Rovelli2007} C. Rovelli, \textit{Quantum Gravity} (Cambridge University Press, Cambridge, England, 2007).
\bibitem{Thiemann2008} T. Thiemann, \textit{Modern Canonical Quantum General Relativity} (Cambridge University Press, Cambridge, England, 2008).

\bibitem{AmelinoCamelia1998}	G. Amelino-Camelia, J. Ellis, N. E. Mavromatos, D. V. Nanopoulos, and S. Sarkar, Nature {\bf393}, 763 (1998).	
\bibitem{Jacob2008}	U. Jacob and T. Piran, JCAP {\bf01}, 031 (2008).	
\bibitem{Abdo2009}	A. A. Abdo et al. (Fermi Collaboration), Nature {\bf462}, 331 (2009).

\bibitem{KosteleckyMewes2002}	V. A. Kostelecky and M. Mewes, Phys. Rev. D {\bf66}, 056005 (2002).
\bibitem{Laurent2011}	P. Laurent et al., Phys. Rev. D {\bf83}, 121301(R) (2011).
\bibitem{Toma2012}	K. Toma et al., Phys. Rev. Lett. {\bf109}, 241104 (2012).

\bibitem{Coleman1999} S. R. Coleman and S. L. Glashow, Phys. Rev. D {\bf59}, 116008 (1999).	
\bibitem{Jacobson2003}	T. Jacobson, S. Liberati, and D. Mattingly, Phys. Rev. D {\bf67}, 124011 (2003).

\bibitem{Igor}
I. Pikovski, M. R. Vanner, M. Aspelmeyer, M. S. Kim, and \v Caslav Brukner, Nature Phys {\bf 8}, 393-397 (2012).
\bibitem{Kumar}
S. P. Kumar, and M. B. Plenio, Nat Commun {\bf 11}, 3900 (2020).



\bibitem{M. Takeda}
M. Takeda, N. Hayashida, et al, Phys. Rev. Lett. {\bf81}, 1163 (1998).

\bibitem{J. Collins}
J. Collins, A. Perez, D. Sudarsky, L. Urrutia, and H. Vucetich, Phys. Rev. Lett. {\bf 93}, 191301 (2004).
\bibitem{J. Polchinski}
J. Polchinski, Classical Quant. Grav. {\bf 29}, 088001 (2012).

\bibitem{Unruh1976} W. G. Unruh, Phys. Rev. D {\bf14}, 870 (1976).
\bibitem{DeWitt1979} B. S. DeWitt, Quantum gravity: The new synthesis, in General Relativity: An Einstein Centenary Survey, edited by S.W. Hawking and W. Israel (Cambridge University Press, Cambridge, England, 1979).
\bibitem{Birrell1982} N. D. Birrell and P. C. W. Davies, Quantum Fields in Curved Space, Cambridge Monographs on Mathematical Physics (Cambridge University Press, Cambridge, England, 1982).
\bibitem{Takagi1986} S. Takagi, Prog. Theor. Phys. Suppl. {\bf88}, 1 (1986).
\bibitem{Crispino2008} L. C. B. Crispino, A. Higuchi, and G. E. A. Matsas, Rev. Mod. Phys. {\bf80}, 787 (2008).
\bibitem{Hu2012} B. L. Hu, S.-Y. Lin, and J. Louko, Classical Quant. Grav. {\bf29}, 224005 (2012).



\bibitem{Louko2008}
J. Louko and A. Satz, Class. Quantum Grav. {\bf 25}, 055012 (2008).
\bibitem{Louko2006}
J. Louko and A. Satz, Class. Quantum Grav. {\bf 23}, 6321 (2006).
\bibitem{Lin2017}
Lin, SY. J. High Energ. Phys. {\bf 2017}, 102 (2017).
\bibitem{Eduardo2013}
E. Mart\'in-Mart\'inez, M. Montero, and M. del Rey, Phys. Rev. D {\bf 87}, 064038 (2013).
\bibitem{Rick2024}
T. Rick Perche, J. Polo-G\'omez, B. de S. L. Terres, and E. Mart\'in-Mart\'inez, Phys. Rev. D {\bf 109}, 045018 (2024).
\bibitem{Keith2018}
Keith K. Ng, R. B. Mann, and E. Mart\'in-Mart\'inez, Phys. Rev. D {\bf 97}, 125011 (2018).
\bibitem{Henderson2018}
L. J. Henderson, R. A. Hennigar, R. B. Mann, A. R. H. Smith, J Zhang, Class. Quantum Grav. {\bf 35} (2018) 21LT02.
\bibitem{Tian2023}
Z. Tian, J. Jing, and J. Du, Sci. China Phys. Mech. Astron. {\bf 66}, 110412 (2023). 



\bibitem{Yu2005} H. Yu, and S. Lu, Phys. Rev. D {\bf72}, 064022 (2005).
\bibitem{Zhu2010} Z. Zhu, and H. Yu, Phys. Rev. A {\bf82}, 042108 (2010).
\bibitem{Arias2016} E. Arias, J. G. Due\~{n}as, G. Menezes, and N. F. Svaiter, J. High Energ.
Phys. {\bf2016}, 147 (2016).
\bibitem{Liu2018} X. Liu, Z. Tian, J. Wang, and J. Jing, Eur. Phys. J. C {\bf78}, 665 (2018).
\bibitem{Liu2021} X. Liu, J. Jing, Z. Tian, and W. Yao, Phys. Rev. D {\bf103}, 125025 (2021).
\bibitem{Liu2025} X. Liu, Z. Tian, and J. Jing, Sci. China-Phys. Mech. Astron. {\bf68}, 100412 (2025).


\bibitem{V. Husain}
V. Husain and J. Louko, Phys. Rev. Lett. {\bf 116}, 061301 (2016).
\bibitem{Nirmalya22016}
N. Kajuri, Class. Quantum Grav. {\bf 33} (2016) 055007.
\bibitem{Louko22018}
J. Louko and S. D. Upton, Phys. Rev. D {\bf 97}, 025008 (2018).
\bibitem{Stargen22017}
D. J. Stargen, N. Kajuri, and L. Sriramkumar, Phys. Rev. D {\bf 96}, 066002 (2017).
\bibitem{Nirmalya22018}
N. Kajuri, G. Sardar, Physics Letters B {\bf 776} (2018) 412–416.
\bibitem{Tian2025}
Y. Wu, and Z. Tian, arXiv:2409.09257v2.


\bibitem{Tian2021}
Z. Tian, and J. Du,  Phys. Rev. D {\bf 103}, 085014 (2021).
\bibitem{Tian2022}
Z. Tian, L. Wu, L. Zhang, J. Jing, and J. Du, Phys. Rev. D {\bf 106}, L061701 (2022).

\bibitem{Tian2026}
Z. Tian, W. Yao, X. Xiao, M. Wang, J. Wang, and J. Jing, Phys. Lett. B {\bf 874} (2026) 140300.



\bibitem{Leggett1980} A. J. Leggett, Progr. Theoret. Phys. Suppl. {\bf69}, 80 (1980).

\bibitem{Streltsov2017}
A. Streltsov, G. Adesso, and M. B. Plenio, Rev. Mod. Phys. {\bf89}, 041003 (2017).

\bibitem{Baumgratz2014}
T. Baumgratz, M. Cramer, and M. B. Plenio, Phys. Rev. Lett. {\bf 113}, 140401 (2014).
\bibitem{Hu2018}
M. L. Hu, X. Hu, J. C. Wang, Y. Peng, Y. R. Zhang, and H. Fan, Phys. Rep. {\bf762}, 1 (2018).



\bibitem{RQI}
R. B. Mann and T. C. Ralph, Class. Quantum Grav. {\bf 29} (2012) 220301.

\bibitem{Wang2016}
Jieci Wang, Zehua Tian, Jiliang Jing, and Heng Fan, Phys. Rev. A {\bf93}, 062105 (2016).

\bibitem{aLiu2016} X. B. Liu, Z. H. Tian, J. C. Wang, and J. L. Jing,
Ann. Phys. {\bf366}, 102-112 (2016).
\bibitem{bLiu2016} X. B. Liu, Z. H. Tian, J. C. Wang, and J. L. Jing,
Quantum Inf. Process. {\bf15}, 3677-3694 (2016).
\bibitem{Liu2022} W. Zhang, X. Liu, and T. Yang, Sci. Rep. {\bf12}, 12577 (2022).



\bibitem{Saveetha2022}
S. Harikrishnan, S. Jambulingam, P. P. Rohde, and C. Radhakrishnan, Phys. Rev. A, {\bf 105}, 052403 (2022).
\bibitem{Du2024}
M. Du, \emph{et al.}, Eur. Phys. J. C (2024) {\bf 84}:838.
\bibitem{Li2024}
H. Li, \emph{et al.}, Eur. Phys. J. C (2024) {\bf 84}:1241.
\bibitem{Wu2021}
S. Wu, H. Zeng, and H. Cao, Class. Quantum Grav. {\bf 38} 185007 (2021).



\bibitem{Wus2023}
S. Wu, C. Wang, D. Liu, X. Huang, and H Zeng, J. High Energ. Phys. {\bf 2023}, 115 (2023).

\bibitem{Wus2024}
W. Li and S. Wu, J. High Energ. Phys. {\bf 2024}, 144 (2024).

\bibitem{Barros2025}
P. H. M. Barros, P. R. S. Carvalho, H. A. S. Costa, Eur. Phys. J. C (2025) {\bf 85}:868.
\bibitem{Barros2024}
Pedro H. M. Barros, Helder A. S. Costa, Eur. Phys. J. C (2024) {\bf 84}:1261.

\bibitem{Nicolaos2023}
N. K. Kollas, D. Moustos, and M. R. Mu\~noz, Phys. Rev. A {\bf 107}, 022420 (2023).

\bibitem{TianSCPMA2023}
Z. Tian, J. Jing, and J. Du, Sci. China Phys. Mech. Astron. {\bf 66}, 110412 (2023).


\bibitem{Enrique2018}
E. Arias, T. R. de Oliveira, and M. S. Sarandy, J. High Energ. Phys. {\bf 02} (2018) 168.

\bibitem{Finnian2018}
F. Gray, and R. B. Mann, J. High Energ. Phys. {\bf 11} (2018) 174.

\bibitem{Arnab2022}
A. Mukherjee, S. Gangopadhyay, and A. S. Majumdar, J. High Energ. Phys. {\bf 09} (2022) 105.

\bibitem{Dimitris2025}
D. Moustos, and O. Abah, Phys. Rev. D {\bf 111}, 105030 (2025).


\bibitem{Vahid2025}
V. Shaghaghi \emph{et al.}, arXiv:2508.20692v1.

\bibitem{Hao2020}
H. Xu, and M. Yung, Phys. Let B {\bf 801} (2020) 135201.

\bibitem{Rudra2025}
R. P. Sarkar, A. Mukherjee, and S. Gangopadhyay,  arXiv:2507.20928v1.

\bibitem{Nikos2024}
N. K. Kollas, and D. Moustos, Phys. Rev. D {\bf 109}, 065025 (2024).

\bibitem{Tomoya2025}
T. Hirotani, and K. G. Yoshimura, Phys. Rev. D {\bf 111}, 125023 (2025).

\bibitem{DimitrisarXiv2025}
D. Moustos, and O. Abah, arXiv:2508.11554v1.


\bibitem{TianJHEP2025}
Z. Tian, X. Liu, J. Wang, and J. Jing, J. High. Energ. Phys. {\bf 2025}, 188 (2025).

\bibitem{LiuSCPMA2025}
X. Liu, Z. Tian, J. Jing, Sci. China Phys. Mech. Astron. {\bf 68}, 100412 (2025).

\bibitem{ArnabarXiv2025}
A. Mukherjee, S. Gangopadhyay, A. S. Majumdar, arXiv:2411.02849v1.

\bibitem{YanPRD2025}
Y. Chen, W. Zhang, T. Ren, and X. Hao, Phys. Rev. D {\bf 111}, 065028 (2025).

\bibitem{ZhiAS2025}
Z. Liu, Z. Tian, and J. Wang, Adv. Sci. {\bf 2025}, e20281 (2025).

\bibitem{RahularXiv2025}
R. Shastri, arXiv:2512.07567v2.

\bibitem{2Husain2010} 
G. M. Hossain, V. Husain, and S. S. Seahra, Phys. Rev. D {\bf82}, 124032 (2010).


\bibitem{velocity}
https://doi.org/10.1142/9789814436403\_0021.

\bibitem{Breuer2002}
H. -P. Breuer and F. Petruuccione, \emph{The Theory of Open Quantum Systems}, (Oxford University Press, Oxford, 2002).










\bibitem{Willis E. Lamb}
Willis E. Lamb and Robert C. Retherford, Phys. Rev. {\bf 72}, 241šC243 (1947).

\bibitem{Ashtekar2004} A. Ashtekar, J. Lewandowski, Class. Quant. Grav. {\bf21}, R53-R152 (2004).

\bibitem{1Ashtekar2003} A. Ashtekar, S. Fairhurst, and J. L. Willis, Classical Quantum Gravity {\bf20}, 1031 (2003).
\bibitem{2Ashtekar2003} A. Ashtekar, J. Lewandowski, and H. Sahlmann, Classical Quantum Gravity {\bf20}, L11 (2003).
\bibitem{1Husain2010} V. Husain and A. Kreienbuehl, Phys. Rev. D {\bf81}, 084043 (2010).
\bibitem{3Husain2010} G. M. Hossain, V. Husain, and S. S. Seahra, Phys. Rev. D {\bf81}, 024005 (2010).
\bibitem{Seahra2012} S. S. Seahra, I. A. Brown, G. M. Hossain, and V. Husain, J. Cosmol. Astropart. Phys. {\bf10}, 041 (2012).

\bibitem{1Husain2016} G. M. Hossain and G. Sardar, Class. Quantum Grav. {\bf33}, 245016 (2016).
\bibitem{Kajuri2016} N. Kajuri, Class. Quantum Grav. {\bf33}, 055007 (2016).
\bibitem{Kajuri2018} N. Kajuri, G. Sardar, Physics Letters B. {\bf776}, 412-416 (2018).
\bibitem{Louko2018} J. Louko and S. D. Upton, Phys. Rev. D {\bf97}, 025008 (2018).

\bibitem{Husain2015} G. M. Hossain and G. Sardar, Phys. Rev. D {\bf92}, 024018 (2015).
\bibitem{Stargen2017} D. J. Stargen, N. Kajuri, and L. Sriramkumar, Phys. Rev. D {\bf96}, 066002 (2017).

\bibitem{A. Garcia-Chung}
A. Garcia-Chung, M. F. Carney, J. B. Mertens, A. Parvizi, S. Rastgoo and Y. S. Tavakoli, JCAP {\bf 11} (2022) 054.

\end{thebibliography}
\end{document}